\documentclass[preprint,review,11.5pt]{elsarticle}

\usepackage{graphicx,dblfloatfix}
\usepackage{makeidx}  
\usepackage{booktabs} 
\usepackage{amsmath}
\usepackage{nccmath}
\usepackage{float}
\usepackage{amsfonts}
\usepackage{comment} 

\usepackage[table,xcdraw]{xcolor}
\usepackage{csquotes}
\usepackage[ruled]{algorithm2e} 

\usepackage{subcaption}
\usepackage{hyperref}
\usepackage{multirow}
\usepackage[normalem]{ulem}
\useunder{\uline}{\ul}{}

\begin{document}

\begin{frontmatter}

\title{Automated Discovery of Business Process Simulation Models from Event Logs}


\author[1,2]{Manuel Camargo\corref{cor1}}
\ead{manuel.camargo@ut.ee}
\author[1]{Marlon Dumas}
\ead{marlon.dumas@ut.ee}
\author[2]{Oscar Gonz\'alez-Rojas}
\ead{o-gonza1@uniandes.edu.co}

\address[1]{University of Tartu, Tartu, Estonia}
\address[2]{Universidad de los Andes, Bogot\'a, Colombia}
\cortext[cor1]{Corresponding author}


\begin{abstract}
Business process simulation is a versatile technique to estimate the performance of a process under multiple scenarios. This, in turn, allows analysts to compare alternative options to improve a business process.
A common roadblock for business process simulation is that constructing accurate simulation models is cumbersome and error-prone. 
Modern information systems store detailed execution logs of the business processes they support. Previous work has shown that these logs can be used to discover simulation models. However, existing methods for log-based discovery of simulation models do not seek to optimize the accuracy of the resulting models. Instead they leave it to the user to manually tune the simulation model to achieve the desired level of accuracy.
This article presents an accuracy-optimized method to discover business process simulation models from execution logs. The method decomposes the problem into a series of steps with associated configuration parameters. A hyper-parameter optimization method is used to search through the space of possible configurations so as to maximize the similarity between the behavior of the simulation model and the behavior observed in the log. The method has been implemented as a tool and evaluated using logs from different domains.
\end{abstract}


\end{frontmatter}

\section{Introduction}
\label{sec:intro}

Business Process Simulation (BPS) is a widely used technique for quantitative analysis of business processes~\cite{FBPM}. 
The main idea of BPS is to generate a set of possible execution traces of a process from a business process model annotated with parameters such as the arrival rate of new process instances, the processing time of each activity, etc. The resulting execution traces are then used to compute performance measures of the process, for example, cycle time, resource utilization, and waiting times for each task in the process.

BPS is commonly used when deciding how to improve a business process with respect to one or more cost and time-related performance measures~\cite{Wynn2008}. For example, an analyst may use BPS to compare the effect of adding two more resources to a process, versus adding only one additional resource, or parallelizing multiple activities that are currently performed sequentially.

A key ingredient for BPS is the availability of a simulation model (herein a BPS model) that accurately reflects the actual dynamics of the process. Traditionally, BPS  models are created by domain experts using manual data gathering techniques, such as interviews, contextual inquiries, and on-site observation. This approach is time-consuming~\cite{Maruster2009}. Furthermore, the accuracy of a BPS model discovered in this way is limited by the accuracy of the business process model that is used as a starting point. Yet, oftentimes process models produced by domain experts do not capture all possible execution paths (e.g.\ exceptional paths are left aside). Indeed, given that business process models are often designed for documentation and communication purposes, they need to strike a balance between completeness and understandability.

Previous studies have advocated the use of Process Mining (PM) techniques to discover BPS models from business process execution logs (also known as \emph{event logs})~\cite{Rozinat2009,Martin2016}. The key idea behind these studies is that a business process simulation model can be obtained by first extracting a process model from an event log using an \emph{automated process discovery} technique, and then enhancing this model with simulation parameters derived from the event log (e.g.\ arrival rate, processing times, and conditional branching probabilities).

However, existing proposals in this area do not consider the question of measuring and automatically tuning the accuracy of the resulting BPS models. Instead, existing proposals leave it to the user to manually tune the BPS model.

This article addresses this gap by proposing an automated method to discover an accuracy-optimized BPS model from an event log. The method decomposes the problem at hand into a series of steps, each of which can be configured via one or more hyper-parameters.\footnote{The term \emph{hyper-parameter} is used to refer to any of the parameters of the method for discovering BPS models, while the term \emph{parameter} refers to any of the parameters of the generated BPS model, such as the arrival rate or the processing times of activities.} A hyper-parameter optimization method is used to maximize the similarity between the behavior generated by the BPS model and the behavior observed in the log, with respect to a similarity measure that takes into account both the ordering of activities and their execution times. 

The proposed method has been implemented as a tool (namely Simod) that generates process simulation models from an event log in the eXtensible Event Stream (XES) format. The resulting simulation model that can be executed using two simulators: BIMP~\cite{BIMP} and Scylla~\cite{Scylla}. The magnitude of the accuracy enhancements achieved by the proposed method have been evaluated via experiments on three event logs from different domains.

This article is a significantly extended version of a tool demonstration paper~\cite{Camargo2019}. The previous tool paper outlines the high-level architecture of Simod. This article adds a detailed description of the algorithms employed, a definition of the BPS model accuracy measure, and an evaluation of the proposed method.


The rest of the article is structured as follows. Section~\ref{sec:background} introduces basic terminology in the field of BPS and discusses existing approaches for BPS model discovery. Section~\ref{sec:tool} presents the proposed method and the approach for measuring BPS model accuracy. Section~\ref{sec:evaluation} discusses the experimental evaluation. Finally, Section~\ref{sec:conclusion} draws conclusions and outlines directions for future work.

%
\section{Background and Related Work}
\label{sec:background}

This section presents background concepts used throughout the article, followed by an analysis of the related work in data-driven BPS.

\subsection{Business process simulation}

This article considers business process models represented in the Business Process Model and Notation (BPMN). 
In its basic form, a BPMN process model consists of activity nodes (or \emph{activities} for short) and gateways that are interconnected by sequence flows. A \emph{split gateway} has multiple outgoing sequence flows. An \emph{exclusive decision gateway} is a split gateway that encodes one decision, i.e.\ when the execution of the process reaches this gateway only one of its outgoing sequences flows is taken. An \emph{inclusive decision gateway} allows multiple of its outgoing flows to be taken when the branching conditions are satisfied. Any inclusive decision gateway can be trivially transformed into a combination of exclusive decision gateways and parallel gateways, hence we can restrict ourselves to exclusive decision gateways without loss of generality. The branches coming out of a decision gateway are called \emph{conditional branches}.

The \emph{cycle time} of a process instance (herein called a \emph{case}) is the amount of time between the moment the case starts its execution and the moment it ends. By extension, we define the cycle time of an instance of an activity as the amount of time between the moment the activity instance is enabled (i.e.\ ready to be executed) and the moment it completes. The \emph{processing time} of an activity instance is the amount of time between the moment the activity instance is started and the moment it is completed. Usually, there is a delay between the moment an activity instance is enabled and the moment it starts. This delay is called \emph{waiting time}. We define the processing time of a case as the amount of time when the process instance is \emph{active}, meaning that at least one activity instance of this case has started but not yet completed. The \emph{waiting time} of a case is the cycle time of the case minus the processing time. 
These definitions also apply to a process, which consists of a set of cases. The cycle time of a process is the mean cycle time of its cases. Similarly, the cycle time of an activity is the mean cycle time of its activity instances.

A BPS model consists of a process model plus the following elements~\cite{FBPM}:
\begin{itemize}
\item The \emph{mean inter-arrival time} of cases and its associated probability distribution function, e.g.\ one case is created every 10 seconds on average with an exponential distribution.
\item The probability distribution of the processing times of each activity. For example, the processing times of an activity may follow a normal distribution with a mean of 20 minutes and a standard deviation of 5 minutes, or an exponential distribution with a mean of 10 minutes.
\item For each conditional branch in the process model, a \emph{branching probability} (i.e.\ percentage of time the conditional branch in question is taken when the corresponding decision gateway is reached.
\item The \emph{resource pool} that is responsible for performing each activity in the process model. For example, in an insurance claims handling process, a possible resource pool would be the \emph{claim handlers}. Each resource pool has a size (e.g., the number of claim handlers or the number of clerks). The instances of a resource pool are the \emph{resources}.
\item A \emph{timetable} for each resource pool, indicating the time periods during which a resource of a resource pool is available to perform activities in the process (e.g.\ Monday-Friday from 9:00 to 17:00).
\item A function that maps each task in the process model to a resource pool.
\end{itemize}

A BPS model consisting of the above elements can be executed using a discrete event simulator. The simulation of a BPS model yields a simulation log, consisting of the execution traces generated during the execution, as well as a collection of performance measures (e.g.\ cycle time, resource utilization, etc.). 


\subsection{Related work}

Data-driven approaches to BPS can be classified in two categories. The first category consists of approaches that provide conceptual guidance to discover BPS models. The second category consists of approaches that seek to automate the discovery of BPS models. Below we review each of these two categories.

\paragraph{\textbf{\textit{Conceptual guidance for data-driven simulation}}}
These approaches discuss how process mining techniques can be used to extract, validate and tune BPS model parameters, without seeking to provide fully automated support. 

Martin et al.~\cite{Martin2014} identify four components of a simulation model, namely entities, activities, resources and gateways. The authors identify BPS modeling tasks related to each of these components (e.g., modeling gateways, modeling activities). In \cite{Martin2016} the same authors present a literature review on the use of process mining techniques to support each of these modeling tasks. This review sheds insights into the question of how to choose process mining techniques for each of the BPS modeling tasks. In this paper, we use these insights as a basis to design an automated method for discovery of BPS models from event logs.


In~\cite{Wynn2008}, the authors present an approach to enhance a given process model with simulation parameters. This approach differs from the one presented in the present article in that it assumes that process model is given as input (in addition to the event log). The approach also assumes that the process model perfectly fits the event log. In reality though, the traces in the event log may deviate with respect to the behavior captured in the process model. Moreover, the approach in~\cite{Wynn2008} does not seek to provide an automated end-to-end approach for discovering BPS models from event logs, but it rather focuses on providing guidance for approaching some of the steps in the discovery of a BPS model.

The authors in \cite{Maruster2009} present a methodology for process improvement based on data-driven simulation. The authors propose as a first step the discovery of simulation models from data for the representation of the current state of the process. Next, the authors propose the manual evaluation of possible scenarios to lead the process to a desired state. The work is illustrated with three case studies from the gas, government and agriculture industries. This work is useful for scoping the relevance of using data for discovering simulation models, as well as for identifying the need for automating this task to explore possible scenarios more quickly and efficiently. However, it does not provide concrete guidance for discovering accurate BPS models from process execution data.

\paragraph{\textbf{\textit{Automated discovery of BPS models using process mining}}}
The methods in this category seek to automate the discovery of BPS models from event logs by means of process mining techniques. 
Rozinat et al.~\cite{Rozinat06miningcpn} propose a semi-automatic approach to discover BPS models based on Colored Petri Nets (CPNs). In this work, an event log is used as input for the discovery of various elements of a BPS model, including the process model, the conditional branching probabilities, and the resource pools. However, the automatic discovery of activity processing times and case inter-arrival times (and their probability distributions) are left aside. 
In \cite{Rozinat2009}, the authors go further by proposing a technique to discover more complete BPS models that include processing times and case inter-arrival times.
These simulation parameters are then combined with the process model into a single CPN, which can be simulated using a CPN tool.
One limitation of the work of Rozinat et al.~\cite{Rozinat2009} is that it does not seek to automatically adjust or fit the probability distributions of the processing times of activities (nor the probability distribution of case inter-arrival times). 
Also, the step where the multiple model BPS model elements are merged together is not automated. Moreover, Rozinat et al. do not seek to optimize the accuracy of the BPS model. The authors suggest to measure the accuracy of the simulation model by comparing the cycle time produced by the BPS model to the ground truth, but this only provides a coarse-grained assessment. Two event logs may have similar cycle times, yet the activities in the corresponding traces may occur at very different points in time and in different order.


Khodyrev et al.~\cite{Khodyrev2014} propose a process mining approach to generate BPS models tailored for short-term prediction of performance measures. The authors extract the structure of the process as a Petri net and establish the dependencies between elements and variables using decision trees.  
A limitation of this approach is that it does not discover the resource perspective (i.e., the resource pools) of the BPS model. Instead, it assumes that an infinite amount of resources may perform each activity. And while the authors automatically discover the conditional branching probabilities and of the activity processing times, the integration of these elements into a BPS model is left to the user. Moreover, the approach does not define how to measure and optimize the accuracy of the resulting BPS model.

Finally, Gawin et al.~\cite{Gawin2015} combine multiple process mining techniques to create a BPS model that reflects the actual process behavior. Specifically, process mining techniques are employed to extract the process model structure, the resource pools, the activity processing times, and the decision logic of decision gates. Interviews and process documentation techniques are used to elicit the case inter-arrival times, the costs of resources use, and the definition of resource schedules. The simulation parameters discovered in this way are then manually linked in the ADONIS tool, leading to a BPS model that is then executed with a capacity analysis algorithm of this latter tool. This latter work differs from the one reported in this article in that it does not seek to automate the extraction of all elements of a BPS simulation model (nor their assembly). Also, it does not seek to measure and optimize the accuracy of the BPS model.


Table \ref{tab:character} summarizes the capabilities of the above approaches for BPS model discovery. In this table, the symbol $(+)$ implies that the feature is supported, $(-)$ implies not supported, and $(+/-)$ implies partially supported, for example, supported but not in an automated manner. 

\begin{table}[ht]
\scriptsize
\centering
\resizebox{\textwidth}{!}{%
\begin{tabular}{@{}l|c|c|c|c@{}}
\toprule
\textbf{Characteristics} & \multicolumn{1}{l|}{\textbf{\begin{tabular}[c]{@{}l@{}}Rozinat et al.\\ (2006)\cite{Rozinat06miningcpn}\end{tabular}}} & \multicolumn{1}{l|}{\textbf{\begin{tabular}[c]{@{}l@{}}Rozinat et al.\\ (2009)\cite{Rozinat2009}\end{tabular}}} & \multicolumn{1}{l|}{\textbf{\begin{tabular}[c]{@{}l@{}}Khodyrev et al.\\ (2014)\cite{Khodyrev2014}\end{tabular}}} & \multicolumn{1}{l}{\textbf{\begin{tabular}[c]{@{}l@{}}Gawin et al.\\ (2015)\cite{Gawin2015}\end{tabular}}} \\ \midrule
\textbf{Sequence flow discovery} & $(+)$ & $(+)$ & $(+)$ & $(+)$ \\ \midrule
\textbf{Resource pools discovery} & $(+)$ & $(+)$ & $(-)$ & $(+)$ \\ \midrule
\textbf{\begin{tabular}[c]{@{}l@{}}Branching probabilities discovery\end{tabular}} & $(+)$ & $(+)$ & $(+/-)$ & $(+)$ \\ \midrule
\textbf{Probabilities distribution fitting} & $(-)$ & $(-)$ & $(-)$ & $(-)$ \\ \midrule
\textbf{Model assembly} & $(-)$ & $(-)$ & $(-)$ & $(-)$ \\ \midrule
\textbf{Accuracy assessment} & $(-)$ & $(+/-)$ & $(+/-)$ & $(+/-)$ \\ \midrule 
\textbf{Accuracy optimization} & $(-)$ & $(-)$ & $(-)$ & $(-)$ \\ \bottomrule
\end{tabular}%
}
\caption{Comparison of approaches to discover and/or enhance BPS models}
\label{tab:character}
\end{table}

This article advances the state-of-the-art in two ways. First, it proposes a fully automated method for discovering each of the perspectives of a BPS model and assembling the resulting perspectives into a complete BPS model. Second, it proposes an approach to measure the accuracy of a BPS model and to optimize the accuracy of an automatically discovered BPS model.  

\section{Approach}
\label{sec:tool}

The proposed method takes as input an event log specified in XES or CSV format, in which every event (corresponding to the execution of an activity instance) has the following attributes: a case identifier, an activity label, a resource which performed the activity, the start timestamp, and the end timestamp\footnote{Alternatively, the event log can also be processed in scenarios where each activity instance is recorded as a start event and an end event, each one with a corresponding timestamp.}. The resource attribute is required to discover the available resource pools, their timetables, and the mapping between activities and resource pools. Equally the start and end timestamps are required to compute the processing time of activities and their respective probability distributions. 

\begin{figure}[ht]
  \begin{center}
    \includegraphics[width=.9\textwidth]{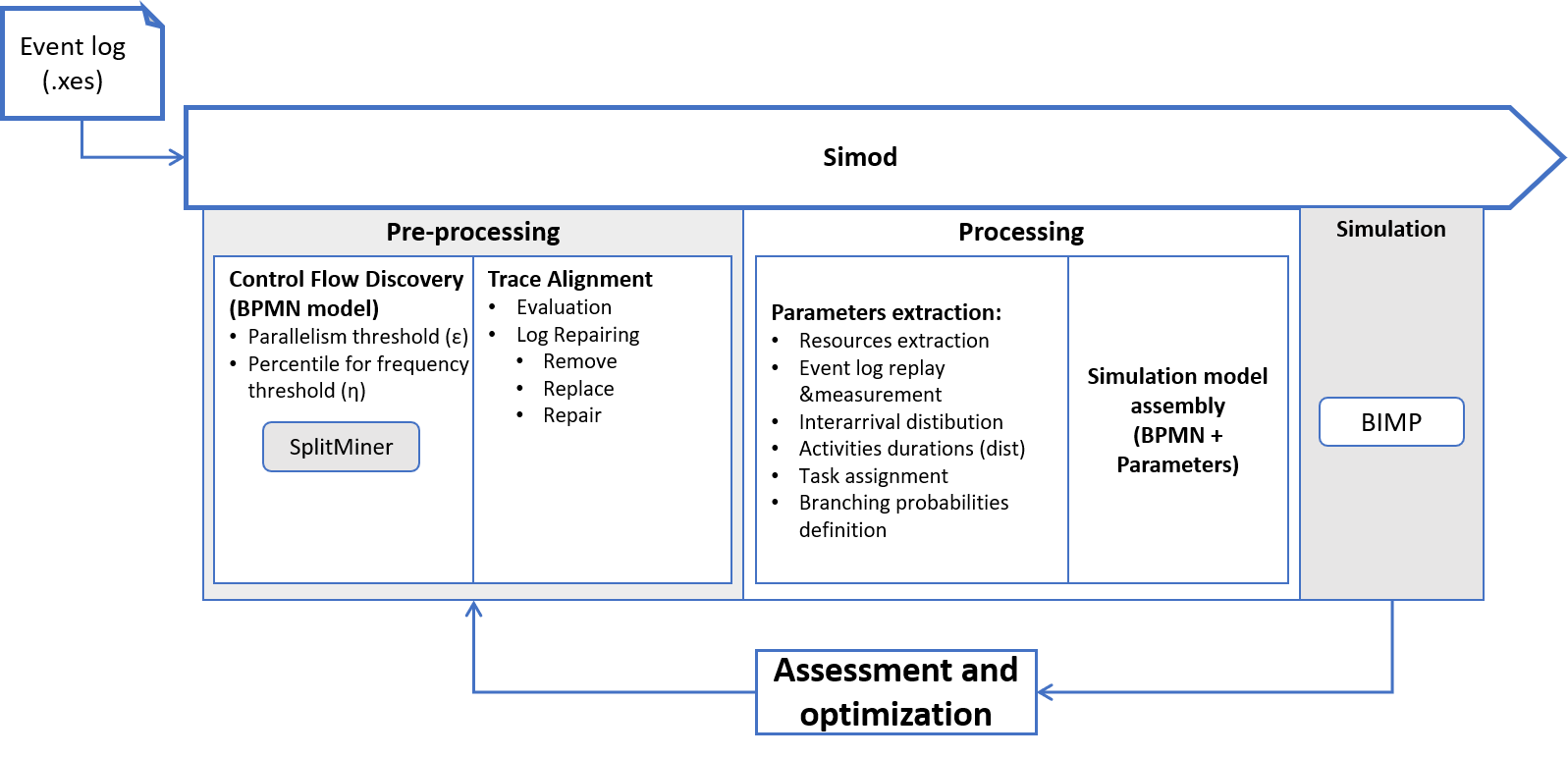}
    \caption{Steps of the BPS model discovery method}
    \label{fig:architecture}
   \end{center}
\end{figure}

Figure \ref{fig:architecture} illustrates the steps of the proposed method for automating the discovery of BPS models. These steps are explained and exemplified in the following subsections by using a synthetic event log of a purchase-to-pay (P2P) process. This event log \footnote{Taken from the tutorial of the Disco process mining tool, available at: \url{http://fluxicon.com/academic/material/}} consists of 21 activities, 27 resources, and 9119 events related to 608 cases.

\subsection{Pre-processing Stage}
\label{sec:preproc}

The steps at this stage allow extracting a BPMN process model from the event log and guarantee their conformance. Most of the time, due to the characteristics of the process discovery algorithms, the models do not reflect all the possible paths in a business process (the fitness is not 100\%). Therefore, 
the proposed method provides the possibility of applying repair actions on the log in order to improve the fitness between the model and the log.

\paragraph{\textbf{Control Flow Discovery}} 
We use the Split Miner algorithm~\cite{Augusto2017} to generate BPMN v2.0 models from event logs. We selected this process discovery method since it achieves high levels of accuracy (precision and fitness) while at the same time producing simple process models
~\cite{augusto2018automated}. However, there is no limitation to use other process discovery methods (e.g. Inductive Miner~\cite{LeemansFA18}). 

\begin{figure*}[ht]
  \begin{center}
    \includegraphics[width=\textwidth]{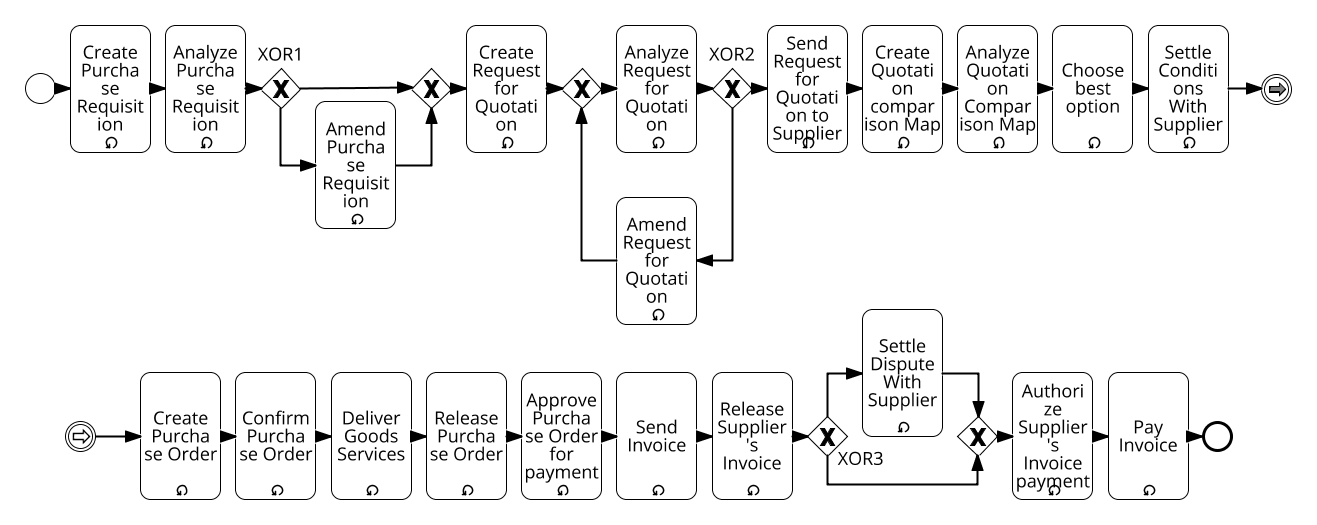}
    \caption{SplitMiner BPMN output example of the purchasing process}
    \label{fig:ex_log}
   \end{center}
\end{figure*}

Split Miner allows the discovery of models with different levels of sensitivity, which depends on the parameters epsilon ($\epsilon$) and eta ($\eta$). The $\epsilon$ parameter refers to the parallelism threshold, which determines the number of concurrent relations captured between events. 
The $\eta$ parameter refers to the percentile for frequency threshold, which only the $\eta$ percentiles most frequent paths between activities. 
Table \ref{tab:ex_log} outlines the structure of the event log used as input, while Figure \ref{fig:ex_log} illustrates the resulting model using $\epsilon$ as 0.3 and $\eta$ as 0.7.

\begin{table}[ht]
\centering
\resizebox{\textwidth}{!}{%
\begin{tabular}{@{}c|c|c|c|c@{}}
\toprule
\textit{\textbf{Case ID}} & \textit{\textbf{Activity}} & \textit{\textbf{Start Timestamp}} & \textit{\textbf{Complete Timestamp}} & \textit{\textbf{Resource}} \\ \midrule
1 & Create Purchase Requisition & 2011/01/01 00:00:00 & 2011/01/01 00:37:00 & Kim Passa \\ \midrule
2 & Create Purchase Requisition & 2011/01/01 00:16:00 & 2011/01/01 00:29:00 & Immanuel Karagianni \\ \midrule
3 & Create Purchase Requisition & 2011/01/01 02:23:00 & 2011/01/01 03:03:00 & Kim Passa \\ \midrule
1 & Create Request for Quotation & 2011/01/01 05:37:00 & 2011/01/01 05:45:00 & Kim Passa \\ \midrule
1 & Analyze Request for Quotation & 2011/01/01 06:41:00 & 2011/01/01 06:55:00 & Karel de G root \\ \midrule
2 & Create Request for Quotation & 2011/01/01 08:16:00 & 2011/01/01 08:26:00 & Alberto Duport \\ \midrule
4 & Create Purchase Requisition & 2011/01/01 08:39:00 & 2011/01/01 09:00:00 & Fjodor Kowalski \\ \bottomrule
\end{tabular}%
}
\caption{Event-log format example}
\label{tab:ex_log}
\end{table}

\paragraph{\textbf{Alignment Evaluation}} 
We measure the degree to which each trace in the log can be aligned with a corresponding trace produced by the process model by using the \emph{fitness measure} proposed in~\cite{AdriansyahDA11}. This alignment is a sequence that has the length of the longest trace and it consists of three symbols: SM (``synchronous move''), MM (``move-on-model'') and ML (``move-on-log''). A SM indicates that the two traces match (i.e.\ the current activity is the same in both traces). A MM means that the two current activities do not match and that the algorithm will ``skip'' the current activity in the model. Thus, the algorithm moves forward in the trace of the model but we stay in the current position in the trace of the log. Conversely, a ML means that the two current activities do not match, thus the current activity is \emph{skipped} in the trace of the log to align the two traces (and remain in the same position in the trace of the model). A perfectly aligned pair of traces contains only SM symbols. Otherwise, the number of MM and ML symbols capture the level of misalignment.

\paragraph{\textbf{Log Repair}}
Once conformance (fitness) is measured, it can be improved by performing a model repair, an event log repair or both \cite{Rogge-SoltiSWMG16}. We perform a log repair for those traces in the log that do not fully fit a trace in the discovered process model. Three methods are proposed for this purpose: removal, replacement, and alignment. 

The \emph{Removal} method omits the traces that are not in conformance with the extracted model, leaving only a reduced log composed of conformant traces. This method is the most natural, and computationally cheap to avoid the outlier traces in the data source. However, if the event log has low conformance with the model, it could end up with a very small event log that could not be sufficiently representative of the process dynamics. 
    
The \emph{Replacement} method replaces each non-conformant trace with a copy of the most similar conformant trace. This action keeps the amount of traces of the log, and globally compensate the lack of alignment between the log and the model. The similarity between traces is defined as one minus the normalized DL distance between two strings. We created an alphabet, by assigning a unique character to each event in the log, to construct words that describe the execution order of the activities. Then each non-conformant trace is compared with all the conformant ones to find the most similar. 

\begin{figure}[ht]
    \centering
    \includegraphics[width=\textwidth]{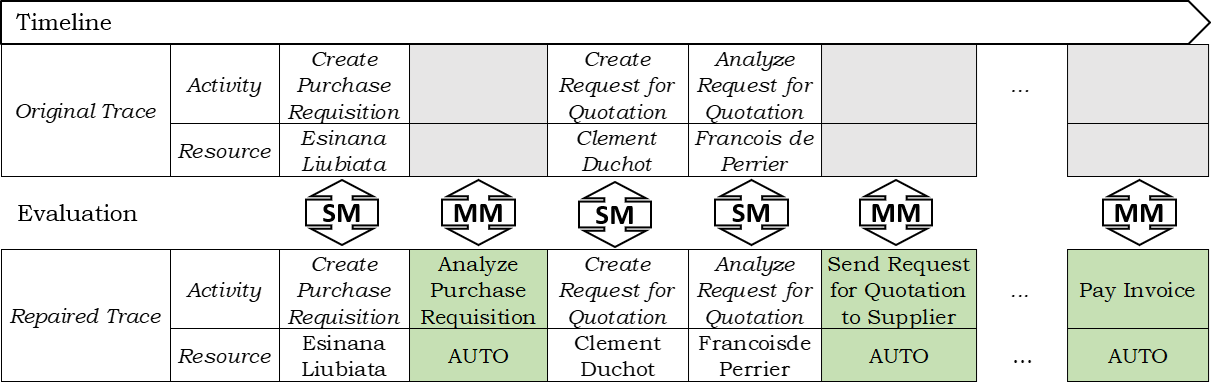}
    \caption{Example: repairing a non-conformant trace}
    \label{fig:pro_analysis}
\end{figure}

The \emph{Alignment} method aligns each process trace of the log with the extracted process model. We use the automata-based alignment technique proposed in the ProConformance 2.0 tool\footnote{http://apromore.org/wp-content/uploads/2017/04/ProConformance2.zip} to determine the optimal alignments, which contain a minimum number of MMs plus MLs in the non-conformant traces. This action keeps the maximum recorded observations preserving the recorded time variability. 
%
To repair a given trace, its corresponding trace alignment is scanned from left to right to apply one of two operations: to remove the event in the trace responsible for an ML, or to annotate the log with zero processing time and a special resource called “AUTO” when a MM is found. This means that this activity does not consume any resources and hence does not have an impact on the cycle time of the process. Finally, the algorithm advances one step in the trace alignment When a SM is found. 
Figure~\ref{fig:pro_analysis} illustrates a trace that was repaired from the original trace with case ID 100 of the aforementioned purchasing process event log. This trace has three activities, ending prematurely with respect to the process model discovered from the log. 
In this case, the ProConformance tool returns a fitness value of 0.4/1 and suggests a type of alignment that can be used to repair the event log: a synchronous move (SM), a move-on-model (MM), or a move-on log (ML).

As an example of the application of the three methods and their differences, suppose we have an event log that has 30\% of traces that do not conform to the discovered model. The Removal technique would eliminate these non-conforming traces, leaving only 70\% for the extraction of parameters. On the other hand, the Replacement technique would look for in the 70\% of conformant traces the most similar to the 30\% non-conformant and duplicate them, resulting in a log of the same size as the original with a 30\% of duplications. Finally, the alignment technique would make adjustments to each of the non-conforming traces, resulting in a log with 70\% of original traces and 30\% of corrected ones.

\subsection{Processing Stage}

At this stage, the tool extracts the simulation parameters and assembles them with the process structure to create a BPS model. 

\paragraph{\textbf{Replay the Event Log}} 
We created a replay algorithm (see Algorithm~\ref{alg:replay}) that takes as input a process model and a repaired trace to calculate the processing time and the enablement time of each activity execution (event) in the trace, as well as the traversal frequency of each conditional flow. These measures are later used to calculate the simulation parameters. 
The \emph{enablement time} is the moment the activity is allowed to start according to the state of the execution. In the simplest case, the enablement time is equal to the end time of the preceding activity in the trace, but this is not always the case, especially in the presence of parallel activities in the process model. 
The \emph{traversal frequency} is the number of times that the conditional branch is traversed while replaying the trace in the log. 
Additionally, the tool calculates the waiting time of each activity by subtracting the start time minus the enablement time. The waiting times calculated in this way are later compared with the waiting times calculated by the simulator to determine the accuracy of the simulation. 
\SetInd{0.7em}{0.1em}
\begin{algorithm}[ht]
\SetAlgoLined
\SetKwInOut{Input}{input}\SetKwInOut{Inputs}{inputs}\SetKwInOut{Output}{output}
\SetKwFor{Foreach}{for each (}{) $\lbrace$}{$\rbrace$}
\tiny
{
    \Inputs{A Process Model M, A trace T}
	\Output{processingTime: A map from events in T to Int }
	\Output{enablementTime: A map from events in T to Timestamp}
	\Output{traversalFrequency: A map from sequence flows in M to Int}
    \Foreach{$e \in T$}
    {
    $processingTime[e] \leftarrow endTime(e) - startTime(e)$\;
        \Repeat{not gatewayFired}{
            $gatewayFired \leftarrow false$\;
            \Foreach{$g \in gateways(M)$}
            {   
                \If{$ isEnabled(M, currentMarking, g)$}
                {
                    \If{$ gatewayType(G) =$ XOR}
                    {
                    $tcf \leftarrow traversedConditionalFlow(M,currentMarking, g, e)$\;
                    $traversalFrequency[tcf]++$\;
                    $currentMarking \leftarrow fire(M, currentMarking, g, e)$\;
                    $gatewayFired \leftarrow true$\;
                    }    
                }    
            }
        }
    \Foreach{$t \in Tasks(M)$ \textbf{where~}$isEnabled(M, currentMarking, g)$}
    {    \If{$  (enablementTime[nextOccurrence(t, T)] \neq \emptyset )$}
        {
          $enablementTime[nextOccurrence(t, T)] \leftarrow currentTime$\;
        }
        
    }
    $currentMarking \leftarrow fire(M, currentMarking, e)$\;
    $currentTime \leftarrow endTime(e)$\;
    }
    \Return $processingTime, enablementTime, traversalFrequency$\;
}
\caption{Replay}
\label{alg:replay}
\end{algorithm}

This algorithm computes the traversal frequencies for each trace in the log, and then sums up them in order to compute the total traversal frequency of each conditional flow. The algorithm relies on the concept of \emph{marking of a BPMN process model}~\cite{Dijkman2008} to capture an execution state with respect to a BPMN model. A marking in a semantically correct (sound) process model is a function that maps each sequence flow in the model to a boolean. A sequence flow is mapped to true if and only if there is a token in that sequence flow in the current state.
The current marking of the model initializes where there is a token in the sequence flow coming out of the start event of the model. Then, the algorithm iterates over each event in the input trace, containing start and end timestamps, to calculate the processing time of the activity. Before handling a given event $e$, the algorithm fires every gateway that is enabled in the current marking, until no more gateways can be fired. When an XOR-gateway is fired, the conditional flow that leads to an activity corresponding to event $e$ is traversed. Accordingly, the traversal frequency of this conditional flow is increased by one.

The algorithm then iterates over the activities enabled in the current marking. If an activity is enabled and the enablement time of the next occurrence of this activity has not yet been initialized (the activity was not enabled before), then the enablement time of this activity is set to be equal to the current execution time. At this point, and given that the model can parse every trace in the repaired input log, the activity corresponding to event $e$ must be enabled. Accordingly, the algorithm fires this activity and updates the current execution time to be equal to the end time of event $e$. The algorithm relies on two auxiliary functions: the $\mathit{isEnabled}$ function determines if a gateway is enabled in the current marking of a BPMN model, whereas the $\mathit{fire}$ function computes the marking reached from the current marking when firing a given gateway. These functions implement the semantics of gateways defined in the BPMN standard. Specifically, a split gateway (with a single incoming flow) is enabled when there is a token in its incoming flow. When it fires, the token is removed from its incoming flow and a token is produced in each of its outgoing flows (in case of an AND gateway) or in one of its outgoing sequence flows (in case of an XOR gateway). In the latter case, the conditional flow leading to enablement of the next event $e$ in the trace is selected. Conversely, an AND-join gateway is enabled if there is a token in each of its incoming flows, while an XOR-join gateway is enabled when there is a token in any of its incoming flows. When a join gateway fires, the tokens in its incoming flows are removed and a token is added to its outgoing flow.

\paragraph{\textbf{Discover the inter-arrival distribution}} 
This step determines the Probability Distribution Function (PDF) of the inter-arrival times for the cases. To this end, the traces in the log are sorted by the start time of their activities. We assume the timestamp of the first event in a trace as the case creation time. Otherwise, a pre-processing task can be used to estimate the actual case creation time as discussed in~\cite{MartinDC15}. 
Then, we calculate the difference between the subsequent start times of the traces daily based. The resulting data series of inter-arrival times are then analyzed to determine which PDF yields the minimum standard error. Our current implementation supports Normal, Exponential, Uniform, Fixed-value, Triangular, Gamma, and Log-normal PDFs. In the running example, we find that the PDF that best fits the observed inter-arrival times is an exponential PDF with a mean of 15455 seconds. 

\begin{figure}[ht]
  \begin{center}
    \includegraphics[width=0.35\textwidth]{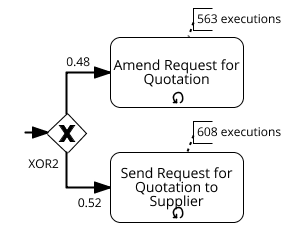}
    \caption{XOR gateway probabilities definition example}
    \label{fig:xor_prob}
    \vspace*{-6mm}
   \end{center}
\end{figure}

\paragraph{\textbf{Conditional branching probabilities}} A BPS model requires defining the probabilities of the paths enabled by decisions made in the gateways. These probabilities can be established by assigning equal values to each conditional branch (e.g. if there are two branches we assign 0.5 probability to each), or by replaying or aligning the traces in the event log against the discovered process model. In the latter case, we normalize the traversal frequencies of the outgoing branches computed during replay, so that their sum is one, hence converting these traversal frequencies into (normalized) probabilities.
In the case of our example event log, the XOR1 gateway has two possible paths to the activities "Amend Request for Quotation" and "Send Request for Quotation to Supplier". These paths were executed in 563 and 608 occasions respectively which means execution probabilities of 0.48 and 0.52 of these paths as is shown in the Figure~\ref{fig:xor_prob}.
\vspace*{-8mm}

\paragraph{\textbf{Activity processing times}} 
We determine the PDF of the processing time of a given activity A in the process model in two steps. First, we create a data series consisting of observed processing time for each execution of activity A in the log (computed by the log replay). Next, we fit a collection of possible distribution functions to the data series to select the distribution function that yields the smallest standard error. For example, we analyze each one of the 21 activities in the purchasing process event log. As can be seen in Table \ref{tab:durdistribution}, most of the processing times follow a Uniform distribution with a mean of 3600 seconds. 

\begin{table}[ht]
\centering
\scriptsize
\resizebox{0.9\textwidth}{!}{%
\begin{tabular}{@{}c|r|r|r|r|r|r|l@{}}
\toprule
\multicolumn{1}{l}{} & \multicolumn{7}{c}{\textbf{PDF}} \\ \midrule
\multicolumn{1}{l|}{} & \multicolumn{1}{c|}{\textbf{Uniform}} & \multicolumn{1}{c|}{\textbf{Normal}} & \multicolumn{1}{c|}{\textbf{Exponential}} & \multicolumn{1}{c|}{\textbf{Gamma}} & \multicolumn{1}{c|}{\textbf{Lognorm}} & \multicolumn{1}{c|}{\textbf{Fixed}} & \multicolumn{1}{c}{\textbf{Triangular}} \\ \bottomrule
\textbf{\# of Activities} & 9 & 2 & -- & 2 & 3 & 5 & -- \\ \midrule
\textbf{Mean} & 3600 & 1285.25 & -- & 1027.55 & 764.66 & 24 & -- \\ \midrule
\textbf{StdDev} & 0 & 136.82 & -- & 548.92 & 704.54 & 32.86 & -- \\ \bottomrule
\end{tabular}%
}
\caption{Purchasing process activities probability distribution functions summary}
\label{tab:durdistribution}
\end{table}

\paragraph{\textbf{Resource pools}} 
Resource pools defining organizational roles and groups are discovered by using the algorithm proposed in~\cite{Song2008}. This algorithm defines activity execution profiles for each resource by creating a graph using the correlation of profiles considering only the relations that overpass a user-defined \emph{similarity threshold}. The resulting graph is a set of unconnected components (clusters) which correspond to groups of resources (resource pool) that generally perform the same type of activities. The algorithm in~\cite{Song2008} then assigns each activity to one or more resource pools. Therefore, we post-process this output to assign each activity to exactly one resource pool that most frequently performs it, as required by a BPS model. In our running example, there are 26 resources that were grouped in 5 resource pools each one assigned to exactly one activity when using the aforementioned algorithm.\footnote{It turns out that this event log contains information about roles and about the mapping from activities to roles. We found that the algorithm in~\cite{Song2008} re-discovered the roles already present in the event log (without using this information) with 100\% accuracy}

\paragraph{\textbf{Simulation model assembly}} Once we have compiled all the simulation parameters, we put them together with the BPMN  model into a single data structure. This step is dependent on the target simulation tool (e.g.\ BIMP or Scylla). In BIMP for example, this step involves embedding the simulation parameters inside the BPMN model, using proprietary XML tags.

\paragraph{\textbf{Simulate Process}} In this last step, the BPS model is given as input to a process model simulator. The simulator outputs a simulated event log. Below we discuss how the accuracy of the resulting BPS model is assessed and optimized.

\subsection{Assessment and optimization}
\label{sec:hyper}

This stage aims to assess the accuracy of the event log generated by the simulator concerning the input event log (i.e., ground truth), and to automatically combine the discovery parameters to obtain the most accurate BPS model. 

\paragraph{\textbf{Assess BPS model accuracy}} 


In order to tune the BPS models produced by Simod, we need to have a way of measuring the accuracy of a BPS model.
We propose to measure accuracy of a BPS model by simulating it and then measuring the similarity (or conversely, the distance) between, on the one hand, the traces in the event log generated by the simulation and, on the other hand, the traces in the log from which the BPS model was discovered (the \emph{ground truth}).\footnote{As an alternative, we could apply a so-called ``holdout method'', where 20\% of traces in the original event log are hold out in a so-called ``testing set''. The remaining 80\% traces of the event log (the "training set") is then used to generate a BPS model using Simod. The accuracy of the resulting BPS model is then assessed using the 20\% of traces in the testing set. This procedure is needed when there is a risk that the BPS model over-fits the event log -- a phenomenon that we did not observe in our experimental evaluation.} To make the log generated by the BPS model comparable to ground truth, we simulate exactly the same number of traces as in the original log, e.g.\ if the original log has 500 traces, we ask the simulator to simulate 500 traces.

To apply the above idea, we need to define a similarity (or a distance) measure between the traces in the simulated log and the traces in the ground truth. One simple way of doing so is by calculating the Mean Absolute Error (MAE) between the cycle times of the simulated traces and of the real traces. While simple to compute, this measure is coarse-grained. Two traces may have the same cycle time, yet consist of very different sets of events.

Another approach to compare pairs of traces is by means of the Damerau-Levenshtein (DL) distance: the minimum number of operations (e.g. adding, deleting, replacing or transposing symbols) required to transform a given string into another. For example, given two traces represented as strings: ``abcd'' and ``acbd'', their DL distance is two. It is easy to normalize this measure so that it returns a number between zero and one, by dividing the absolute DL distance by the maximum length of the two traces. 




The DL distance captures the differences in the activity occurrences and in the ordering of activities. However, it does not take into account two requirements that arise when comparing two business process execution traces
\begin{enumerate}
\item 
The DL distance penalizes transposed occurrences of activities, even if these activities are parallel activities and hence may complete in any order. For example, if activities \textit{b} and \textit{c} are parallel activities, the distance between traces ``abcd'' and ``acbd'' should be zero. The difference is accidental: in one trace \textit{b} occurs before \textit{c} but it could have been vice-versa. 
\item The DL distance does not take into account the waiting times and processing times of the activity occurrences represented by the the events in a trace. Yet, the ability to faithfully capture the waiting times and processing times of activities is a key requirement in business process simulation.
\end{enumerate}

To address the first requirement, we propose to modify the cost function used in the DL distance so that, if two activities are parallel activities, we do not penalize transposed occurrences of these activities in the compared traces.
To this end, we first analyze the input event log in order to discover pairs of parallel activities using the so-called \emph{alpha concurrency oracle}. The alpha concurrency oracle states that two activities (a, b) in an event log are in parallel if sometimes b directly follows a and sometimes a directly follows b. This heuristics is used by the alpha algorithm for process discovery~\cite{VanDerAalst2004a}. Note that this heuristics is not fail-proof. More fine-grained concurrency oracles have been proposed in the literature~\cite{Armas-Cervantes19}, but they are more complex to calculate and they are not fail-proof either. In other words, we could refine this idea by using more fine-grained concurrency oracles at the price of higher computational cost.

Given the concurrency relation $\parallel$ between activities returned by the alpha concurrency oracle, we modify the cost function used in the DL distance so that an occurrence of an activity \emph{b} can be replaced by an occurrence of a parallel activity \emph{c} without penalty, provided that both activities \emph{b} and \emph{c} occur in both input traces. In other words, if these activities co-occur in both input traces, their occurrences are interchangeable.


To address the second requirement, we draw inspiration from~\cite{Dobrisek2009}, which defined a variant of the DL distance for timed words. The idea is that if two events in the input traces have the same activity label, but their waiting and processing times do not match, we assign a penalty when matching this pair of events proportional to the difference between their timestamps. This penalty is normalized so that it is between zero and one.


Based on the above ideas, we propose a modified version of the DL distance, namely the \emph{Business Process Trace Distance} (BPTD). To define the BPTD measure, we first introduce some notations.
We define an event as a tuple $e = (l, p, w)$, where $l$ is a symbol taken from the alphabet of all possible activity labels $L$, $p$ is the processing time and $w$ the event waiting time of the activities -- $p, w \in \mathbb{R}+$. Moreover, let $\mathcal{E}$ is the set of all possible events, i.e.
\begin{ceqn}
\begin{align}
\mathcal{E}=\left\{\left(l,p,w\right)\middle|\ l\in L;\ p, w\in\mathbb{R}+\right\}
\end{align}
\end{ceqn}

A trace is a non-empty sequence of events $\sigma=\langle e_1,e_2,\dots,e_n \rangle$ such that  $e_i=\left(l_i,p_i,w_i\right)\in\mathcal{E},1\le i\le n$. The set of all process traces is $\mathcal{S}$. An event log $\mathcal{L}$ is a set of traces from $\mathcal{S}$ and $\mathcal{K}$ is the number of traces in the event log.  
\begin{ceqn}
\begin{align}
\mathcal{L} = \left\{\sigma_i\middle|\sigma_i \in\mathcal{S}, 1\le i\le\mathcal{K}\right\}
\end{align}
\end{ceqn}

Given two traces $\sigma, \sigma' \in\mathcal{S}$, 
the DL distance between $\sigma$ and $\sigma'$ is the output of following recursive function when initially invoked with $i = \left|\sigma\right|$ and $j= \left|\sigma'\right|$~:

\begin{ceqn}
\scriptsize
\begin{align}
d\left(\sigma, \sigma', i,j\right)=min\left\{
\begin{array}{ll}
0&if\ i=j=0\\
d\left(\sigma, \sigma', i-1,j\right)+c\left(\sigma\langle i\rangle,\sigma^\prime\langle j\rangle\right)&if\ i\ >\ 0\\
d\left(\sigma, \sigma', i,j-1\right)+c\left(\sigma\langle i\rangle,\sigma^\prime\langle j\rangle\right)&if\ j\ >\ 0\\
d\left(\sigma, \sigma', i-1,j-1\right)+c\left(\sigma\langle i\rangle,\sigma^\prime\langle j\rangle\right)&if\ i,j\ >\ 0\\
d\left(\sigma, \sigma', i-2,j-2\right)+c\left(\sigma\langle i\rangle,\sigma^\prime\langle j\rangle\right)&if\ i,j >1\\
&\&\ \sigma\langle i\rangle=\sigma^\prime\langle j-1\rangle\\
& \&\ \sigma\langle i-1\rangle=\sigma^\prime\langle j\rangle\\
\end{array}
\right.
\end{align}
\end{ceqn}

In the classical definition of the DL distance, the cost function $c$ returns one for each deletion, insertion, replacement (mismatch) or transposition. BPTD is defined in the same way as the DL distance, but it uses a cost function the (i) does not penalize the replacement of one activity by another parallel activities if both of these activities co-occur in the input traces; and (ii) introduces a penalty in case two events match (i.e.\ they have same activity label or correspond to parallel activities) but they have different waiting times or processing times. Formally, the cost function used by BPTD is the following one.

\begin{ceqn}
\begin{align}
c(e,e^\prime)=\left\{
\begin{array}{ll} 
\beta\lvert p-p^\prime\lvert + \left(1- \beta\right)\lvert w-w^\prime\lvert &\mathrm{if}\ l=l^\prime \lor (l \parallel l^\prime \wedge l, l' \in \sigma \wedge l, l' \in \sigma') \\
1&otherwise
\end{array}
\right.
\end{align}
\end{ceqn}
\noindent ...where $\lvert p-p^\prime\lvert$ is the absolute error of the normalized processing time and $\lvert w-w^\prime\lvert$ is the absolute error of the normalized waiting time. The coefficient $\beta$ represents the weight given to the processing time and 1 - $\beta$ the weight given to the waiting time (we take $\beta = 0.5$ by default).


To illustrate how the BPTD measures allows us to capture differences in the timing of activities in two traces, let us consider the following example.
Suppose we have two traces $\sigma=\langle e_1,e_2,e_3 \rangle$ and $\sigma^\prime=\langle e^\prime_1,e^\prime_2,e^\prime_3,e^\prime_4 \rangle$ (see Table \ref{tab:metric_example}), and we want to calculate the distance between the two $c(\sigma, \sigma^\prime)$. In the first iteration, the algorithm compares $e_1$ with $e^\prime_1$ finding identical symbols, so it evaluates the time differences. The total time of $e_1$ is 7 for which $\beta = 0.42$, and the cost of the operation is $c(e_1,e^\prime_1) = (0.42\times\lvert0.3-0.2\lvert) +
(0.58\times\lvert0.4-0.4\lvert) = 0.042$. In the next iteration, when comparing $e_2$ with $e^\prime_2$ the symbols are different, but it is possible to transpose $e_2$ and $e_3$. Assuming that the symbols of events $e_2$ and $e^\prime_2$ have a concurrence relationship, the cost of the operation is equal to $c(e_3,e^\prime_2) = (0.8\times\lvert0.4-0.5\lvert) + (0.2\times \lvert0.2-0.1\lvert) = 0.1$, otherwise, the operation cost is 1. In the next iteration after the transposition is made, $e_2$ will be compared again but this time with $e^\prime_3$ in this case the symbols are equal and the times also for which the cost of the operation is $c(e_2, e^\prime_3 ) = (0.83\times\lvert0.5-0.5\lvert) + (0.17\times\lvert0.1-0.1\lvert) = 0$. Finally, an insertion operation is performed to obtain the same traces; the cost of this operation is 1. The total edit cost is 1.142 with concurrency between $e_2$ and $e^\prime_2$, and of 2.042 without it. This value can be normalized by dividing it by the size of the longest trace, as one would do to normalize the DL distance.

\begin{table}[ht]
\centering
\resizebox{\textwidth}{!}{%
\begin{tabular}{@{}llllllcc@{}}
\toprule
\textbf{Iter.} & \textbf{Sec.} & \multicolumn{4}{c}{\textbf{Events}} & \multicolumn{1}{l}{\textbf{\begin{tabular}[c]{@{}l@{}}Cost with\\ concurrency\end{tabular}}} & \multicolumn{1}{l}{\textbf{\begin{tabular}[c]{@{}l@{}}Cost without\\ concurrency\end{tabular}}} \\ \midrule
\multicolumn{1}{|l|}{} & \multicolumn{1}{l|}{$\sigma$} & \multicolumn{1}{l|}{{\color[HTML]{002B36} $e_1 = \left(a, 0.3, 0.4\right)$}} & \multicolumn{1}{l|}{$e_2 = \left(b, 0.5, 0.1\right)$} & \multicolumn{1}{l|}{$e_3 = \left(c, 0.4, 0.1\right)$} & \multicolumn{1}{l|}{} & \multicolumn{1}{c|}{} & \multicolumn{1}{c|}{} \\ \cmidrule(lr){2-6}
\multicolumn{1}{|l|}{\multirow{-2}{*}{0}} & \multicolumn{1}{l|}{$\sigma^\prime$} & \multicolumn{1}{l|}{{\color[HTML]{002B36} $e^\prime_1 = \left(a, 0.2, 0.4\right)$}} & \multicolumn{1}{l|}{$e^\prime_2 = \left(c, 0.5, 0.2\right)$} & \multicolumn{1}{l|}{$e^\prime_3 = \left(b, 0.5, 0.1\right)$} & \multicolumn{1}{l|}{$e^\prime_4 = \left(d, 0.1, 0.1\right)$} & \multicolumn{1}{c|}{\multirow{-2}{*}{0}} & \multicolumn{1}{c|}{\multirow{-2}{*}{0}} \\ \midrule
\multicolumn{1}{|l|}{} & \multicolumn{1}{l|}{$\sigma$} & \multicolumn{1}{l|}{{\color[HTML]{FE0000} \textbf{$e_1 = \left(a, 0.3, 0.4\right)$}}} & \multicolumn{1}{l|}{$e_2 = \left(b, 0.5, 0.1\right)$} & \multicolumn{1}{l|}{$e_3 = \left(c, 0.4, 0.1\right)$} & \multicolumn{1}{l|}{} & \multicolumn{1}{c|}{} & \multicolumn{1}{c|}{} \\ \cmidrule(lr){2-6}
\multicolumn{1}{|l|}{\multirow{-2}{*}{1}} & \multicolumn{1}{l|}{$\sigma^\prime$} & \multicolumn{1}{l|}{{\color[HTML]{FE0000} \textbf{$e^\prime_1 = \left(a, 0.2, 0.4\right)$}}} & \multicolumn{1}{l|}{$e^\prime_2 = \left(c, 0.5, 0.2\right)$} & \multicolumn{1}{l|}{$e^\prime_3 = \left(b, 0.5, 0.1\right)$} & \multicolumn{1}{l|}{$e^\prime_4 = \left(d, 0.1, 0.1\right)$} & \multicolumn{1}{c|}{\multirow{-2}{*}{0.042}} & \multicolumn{1}{c|}{\multirow{-2}{*}{0.042}} \\ \midrule
\multicolumn{1}{|l|}{} & \multicolumn{1}{l|}{$\sigma$} & \multicolumn{1}{l|}{{\color[HTML]{009901} \textbf{$e_1 = \left(a, 0.3, 0.4\right)$}}} & \multicolumn{1}{l|}{{\color[HTML]{FE0000} \textbf{$e_3 = \left(c, 0.4, 0.1\right)$}}} & \multicolumn{1}{l|}{{\color[HTML]{009901} \textbf{$e_2 = \left(b, 0.5, 0.1\right)$}}} & \multicolumn{1}{l|}{} & \multicolumn{1}{c|}{} & \multicolumn{1}{c|}{} \\ \cmidrule(lr){2-6}
\multicolumn{1}{|l|}{\multirow{-2}{*}{2}} & \multicolumn{1}{l|}{$\sigma^\prime$} & \multicolumn{1}{l|}{{\color[HTML]{009901} \textbf{$e^\prime_1 = \left(a, 0.2, 0.4\right)$}}} & \multicolumn{1}{l|}{{\color[HTML]{FE0000} \textbf{$e^\prime_2 = \left(c, 0.5, 0.2\right)$}}} & \multicolumn{1}{l|}{$e^\prime_3 = \left(b, 0.5, 0.1\right)$} & \multicolumn{1}{l|}{$e^\prime_4 = \left(d, 0.1, 0.1\right)$} & \multicolumn{1}{c|}{\multirow{-2}{*}{0.1}} & \multicolumn{1}{c|}{\multirow{-2}{*}{1}} \\ \midrule
\multicolumn{1}{|l|}{} & \multicolumn{1}{l|}{$\sigma$} & \multicolumn{1}{l|}{{\color[HTML]{009901} \textbf{$e_1 = \left(a, 0.3, 0.4\right)$}}} & \multicolumn{1}{l|}{{\color[HTML]{009901} \textbf{$e_3 = \left(c, 0.4, 0.1\right)$}}} & \multicolumn{1}{l|}{{\color[HTML]{FE0000} \textbf{$e_2 = \left(b, 0.5, 0.1\right)$}}} & \multicolumn{1}{l|}{} & \multicolumn{1}{c|}{} & \multicolumn{1}{c|}{} \\ \cmidrule(lr){2-6}
\multicolumn{1}{|l|}{\multirow{-2}{*}{3}} & \multicolumn{1}{l|}{$\sigma^\prime$} & \multicolumn{1}{l|}{{\color[HTML]{009901} \textbf{$e^\prime_1 = \left(a, 0.2, 0.4\right)$}}} & \multicolumn{1}{l|}{{\color[HTML]{009901} \textbf{$e^\prime_2 = \left(c, 0.5, 0.2\right)$}}} & \multicolumn{1}{l|}{{\color[HTML]{FE0000} \textbf{$e^\prime_3 = \left(b, 0.5, 0.1\right)$}}} & \multicolumn{1}{l|}{$e^\prime_4 = \left(d, 0.1, 0.1\right)$} & \multicolumn{1}{c|}{\multirow{-2}{*}{0}} & \multicolumn{1}{c|}{\multirow{-2}{*}{0}} \\ \midrule
\multicolumn{1}{|l|}{} & \multicolumn{1}{l|}{$\sigma$} & \multicolumn{1}{l|}{{\color[HTML]{009901} \textbf{$e_1 = \left(a, 0.3, 0.4\right)$}}} & \multicolumn{1}{l|}{{\color[HTML]{009901} \textbf{$e_3 = \left(c, 0.4, 0.1\right)$}}} & \multicolumn{1}{l|}{{\color[HTML]{009901} \textbf{$e_2 = \left(b, 0.5, 0.1\right)$}}} & \multicolumn{1}{l|}{{\color[HTML]{FE0000} \textbf{$e_4 = \left(d, 0.1, 0.1\right)$}}} & \multicolumn{1}{c|}{} & \multicolumn{1}{c|}{} \\ \cmidrule(lr){2-6}
\multicolumn{1}{|l|}{\multirow{-2}{*}{4}} & \multicolumn{1}{l|}{$\sigma^\prime$} & \multicolumn{1}{l|}{{\color[HTML]{009901} \textbf{$e^\prime_1 = \left(a, 0.2, 0.4\right)$}}} & \multicolumn{1}{l|}{{\color[HTML]{009901} \textbf{$e^\prime_2 = \left(c, 0.5, 0.2\right)$}}} & \multicolumn{1}{l|}{{\color[HTML]{009901} \textbf{$e^\prime_3 = \left(b, 0.5, 0.1\right)$}}} & \multicolumn{1}{l|}{{\color[HTML]{FE0000} \textbf{$e^\prime_4 = \left(d, 0.1, 0.1\right)$}}} & \multicolumn{1}{c|}{\multirow{-2}{*}{1}} & \multicolumn{1}{c|}{\multirow{-2}{*}{1}} \\ \midrule
 & \multicolumn{5}{r}{\textbf{Total cost}} & \multicolumn{1}{l}{\textbf{1.142}} & \multicolumn{1}{l}{\textbf{2.042}} \\ \bottomrule
\end{tabular}%
}
\caption{Exemplification of BPTD measure}
\label{tab:metric_example}
\end{table}

The BPTD measure allows us to compare two traces. But the problem we initially posed was that of comparing two logs . To compare two event logs (the simulated log against the ground-truth log) we define a similarity measure, namely \emph{Event Log Similarity (ELS)}, by pairing each trace in one log with a trace in the other log. Specifically, we search for the pairing of traces that minimizes the sum of the BPTDs between the paired traces. We map the problem of pairing the traces of the two logs to the \emph{assignment problem}, and we use the well-known Hungarian algorithm\cite{Kuhn1955} to find the minimal-distance pairing. Given two logs $L_1$ and $L_2$ and given a minimal-distance pairing $P = \{ (\sigma_1, \sigma_2) \mid \sigma_1 \in L_1 and \sigma_2 \in L_2 \}$ between the traces in these logs:

\[ \mathit{ELS}(\mathit{L}_1, \mathit{L}_2) = \Sigma_{(t_1, t_2) \in P} \mathit{BPTD}(\sigma_1, \sigma_2) \]

For the optimization and evaluation phases, we use the \emph{ELS} similarity function with $\mathit{L}_1$ being the simulated log and $\mathit{L}_2$ being the ground truth.

\paragraph{\textbf{Hyper-parameter optimization}} 
The previous subsections described the automatic creation of a BPS model that integrates multiple perspectives of the process. These perspectives are discovered by using one or more process mining algorithms. However, in each step, critical decisions must be made, either on the techniques to be used or on the parameters' values. For example, the use of low filter values in the BPMN miner can drastically affect the simulation accuracy creating spaghetti models impossible to reproduce by the simulator. 
The same can happen when choosing how similar the resources of the discovered pools should be or when it is required to select the best way to calculate the probabilities of the decision gateways.

In a traditional approach, an expert would manually perform the search of the best combinations by modifying the values of the parameters based on its expertise and intuition. However, this is a time-consuming approach that often leads to far-from-optimal results~\cite{Witt2017}. Therefore, we propose to use a Tree-structured Parzen Estimator (TPE) as a  hyper-parameter optimizer~\cite{Bergstra2011} to find the best settings based on historical accuracy the executed models. TPE is a sequential algorithm that defines on each trial the following parameter configuration. This is based on past results and nested functions that select the parameters' values based on a probability distribution and ranges specified for each one. The objective function seeks to minimize the loss which is calculated as the inverse TSD measure. Table \ref{tab:search_space} defines the used search space. 

\begin{table}[ht]
\centering
\scriptsize
\resizebox{\textwidth}{!}{%
\begin{tabular}{@{}l|l|l|l@{}}
\toprule
\textbf{Category} & \textbf{Variable} & \textbf{Distribution} & \textbf{Range} \\ \midrule
\multirow{2}{*}{Control flow discovery} & Parallelism threshold ($\epsilon$) & Uniform & {[}0...1{]} \\
 & Percentile for frequency threshold ($\eta$) & Uniform & {[}0...1{]} \\ \midrule
\multirow{3}{*}{Log repair technique} & Repair & n/a &  \\
 & Removal & n/a &  \\
 & Replace & n/a &  \\ \midrule
\multirow{3}{*}{Conditional branching probabilities} & Random & n/a &  \\
 & Equiprobable & n/a &  \\
 & Discovered & n/a &  \\ \midrule
Resource pools & Similarity threshold & Uniform & {[}0...1{]} \\ \bottomrule
\end{tabular}%
}
\caption{Search space definition}
\label{tab:search_space}
\vspace*{-6mm}
\end{table}
%
%
\section{Evaluation}
\label{sec:evaluation}

The proposed method has been implemented as an open-source tool, namely Simod, which is packaged as a Python Jupiter Notebook.\footnote{Tool and datasets available at \url{https://github.com/AdaptiveBProcess/Simod}}
Simod takes as input an event log in XES format and produces a BPS model ready to be simulated using the BIMP  simulator~\cite{BIMP}\footnote{Available at \url{http://bimp.cs.ut.ee}}.
The source code of Simod can also be configured to produce models for the Scylla simulator~\cite{Scylla}, but BIMP is used as the default simulator because Scylla only supports a restricted set of probability distributions, thus restricting the space of configuration options.

Using Simod, we conducted an experimental evaluation aimed at addressing the following research questions: (RQ1) What is the accuracy of the BPS models generated by the proposed method? and (RQ2) To what extent the hyper-parameter optimization step improves the accuracy of the BPS models?

\subsection{Datasets}

A pre-requisite to discover a BPS model from an event log, is that the events in the log should have both a start and end timestamps. Unfortunately, this pre-requisite is not fulfilled by publicly available real-life logs such as those in the 4TU Collection of event logs.\footnote{\url{https://data.4tu.nl/repository/collection:event_logs_real}}
As an alternative, we validate the proposed approach using one synthetic event log and two real-life event logs that satisfy the above requirement. The characteristics of these logs are given in Table \ref{tab:event_logs}.
\begin{table}[ht]
\centering
\scriptsize
\resizebox{\textwidth}{!}{%
\begin{tabular}{@{}l|c|c|c|c|c|c|c@{}}
\toprule
\textbf{Event log} & \multicolumn{1}{l|}{\textbf{\begin{tabular}[c]{@{}l@{}}Num. \\ traces\end{tabular}}} & \multicolumn{1}{l|}{\textbf{\begin{tabular}[c]{@{}l@{}}Num.\\ events\end{tabular}}} & \multicolumn{1}{l|}{\textbf{\begin{tabular}[c]{@{}l@{}}Num.\\ activities\end{tabular}}} & \multicolumn{1}{l|}{\textbf{\begin{tabular}[c]{@{}l@{}}Avg.\\ activities\\ per trace\end{tabular}}} & \multicolumn{1}{l|}{\textbf{\begin{tabular}[c]{@{}l@{}}Max.\\ activities\\ per trace\end{tabular}}} & \multicolumn{1}{l|}{\textbf{\begin{tabular}[c]{@{}l@{}}Mean\\ duration\end{tabular}}} & \multicolumn{1}{l}{\textbf{\begin{tabular}[c]{@{}l@{}}Max.\\ duration\end{tabular}}} \\ \midrule
P2P & 608.0 & 9119.0 & 21.0 & 14.9 & 44.0 & 21.5 days & 108 days 7 hours \\ \midrule
ACR & 954.0 & 6870.0 & 18.0 & 7.2 & 23.0 & 14.9 days & 135 days 19 hours \\ \midrule
MP & 225.0 & 4953.0 & 26.0 & 22.0 & 177.0 & 20.6 days & 87 days 10 hours \\ \bottomrule
\end{tabular}%
}
\caption{Statistics of the event logs}
\label{tab:event_logs}
\end{table}

The first event log is a synthetic log, generated from a model not available to the authors, of a purchase-to-pay (P2P) process. This is the same log used as a running example in Section~\ref{sec:tool}.\footnote{The log is part of the academic material of the Fluxicon Disco tool -- https://fluxicon.com/}
The second log stems from an Academic Credentials Recognition (ACR) process at University of Los Andes in Colombia. The log comes from a deployment of a Business Process Management System (BPMS), specifically Bizagi. The model corresponding to this log was not available to the authors of this article. 
The third log is that of a manufacturing production (MP) process, exported from an Enterprise Resource Planning (ERP) system~\cite{production_log}. The tasks in this process refer to steps (or ``stations';) in the manufacturing process.

These three event logs were chosen because they have distinct characteristics in terms of control-flow and they come from different domains. The ACR event log is the one that contains the greatest number of traces and the least number of average activities per trace, while MP is the log with the least traces and the highest number of average activities per trace. On the other hand, the P2P log corresponds to the scenario in which the event log is generated from a process model defined in ideal conditions, so it is expected a 100\% fit between the two, and high accuracy in the simulation. The ACR log corresponds to a service process executed on a BPMS. It is a relatively complex process, which delivers a service to hundreds of users and involves over a dozen workers. Finally, the MP event log is the scenario in which the process structure is unknown, and where the behavior of the resources can affect the structure of the process significantly.

\subsection{Experimental setup}

To address the research questions, we compared a baseline variant of our method against the hyper-parameter optimized variant. The  hyperparameter values used for the baseline are given in the Baseline column in Table~\ref{tab:conf_values}. The hyperparameter-optimized variant explores 100 hyperparameter combinations per log using the TPE optimization method mentioned above. The ranges of hyperparameter values given to the TPE optimizer are shown in Table~\ref{tab:conf_values}.
\begin{table}[ht]
\centering
\resizebox{\textwidth}{!}{%
\begin{tabular}{@{}l|c|c@{}}
\toprule
\multicolumn{1}{c|}{\textbf{Parameter}} & \textbf{Baseline values} & \textbf{Optimizer ranges} \\ \midrule
Parallelism threshold ($\epsilon$) & 0.1 & {[}0...1{]} \\ \midrule
Percentile for frequency threshold ($\eta$) & 0.4 & {[}0...1{]} \\ \midrule
Log repair technique & Removal & Repair, Removal, Replace \\ \midrule
Conditional branching probabilities & Equiprobable & Random, Equiprobable, Discovered \\ \midrule
Similarity threshold & 0.5 & {[}0...1{]} \\ \midrule
Inter-arrival times & Exponential PDF & PDF Discovered from data \\ \midrule
Processing times & Exponential PDF & PDF Discovered from data \\ \bottomrule
\end{tabular}%
}
\caption{Values of the configuration scenarios.}
\label{tab:conf_values}
\end{table}

We discovered BPS models from each event log using both the baseline and the optimized method. We then simulated each BPS model ten times and each time, we measured the similarity between the simulated log and the ground-truth using the ELS measure introduced earlier. A total of 3000 simulated event logs were generated: 1000 per input event log (10 simulation runs for each of 100 parameter combinations tested by the hyper-parameter optimizer).

The results of a simulation may vary from one run to another due to its stochastic nature. To ensure that these stochastic variations are not responsible for the conclusions drawn from the experiments, we applied the single-queue Mann-Whitney U test to validate the significance of the optimization accuracy improvement. The alternative hypothesis was that the accuracy of the best configuration is higher than the accuracy of the baseline scenario. In this evaluation, we compared the ten simulation runs of the best BPS found by the optimizer against the ten simulation runs of the baseline model.

\subsection{Results}

\begin{figure}[!ht]
    \centering
    \includegraphics[width=\textwidth]{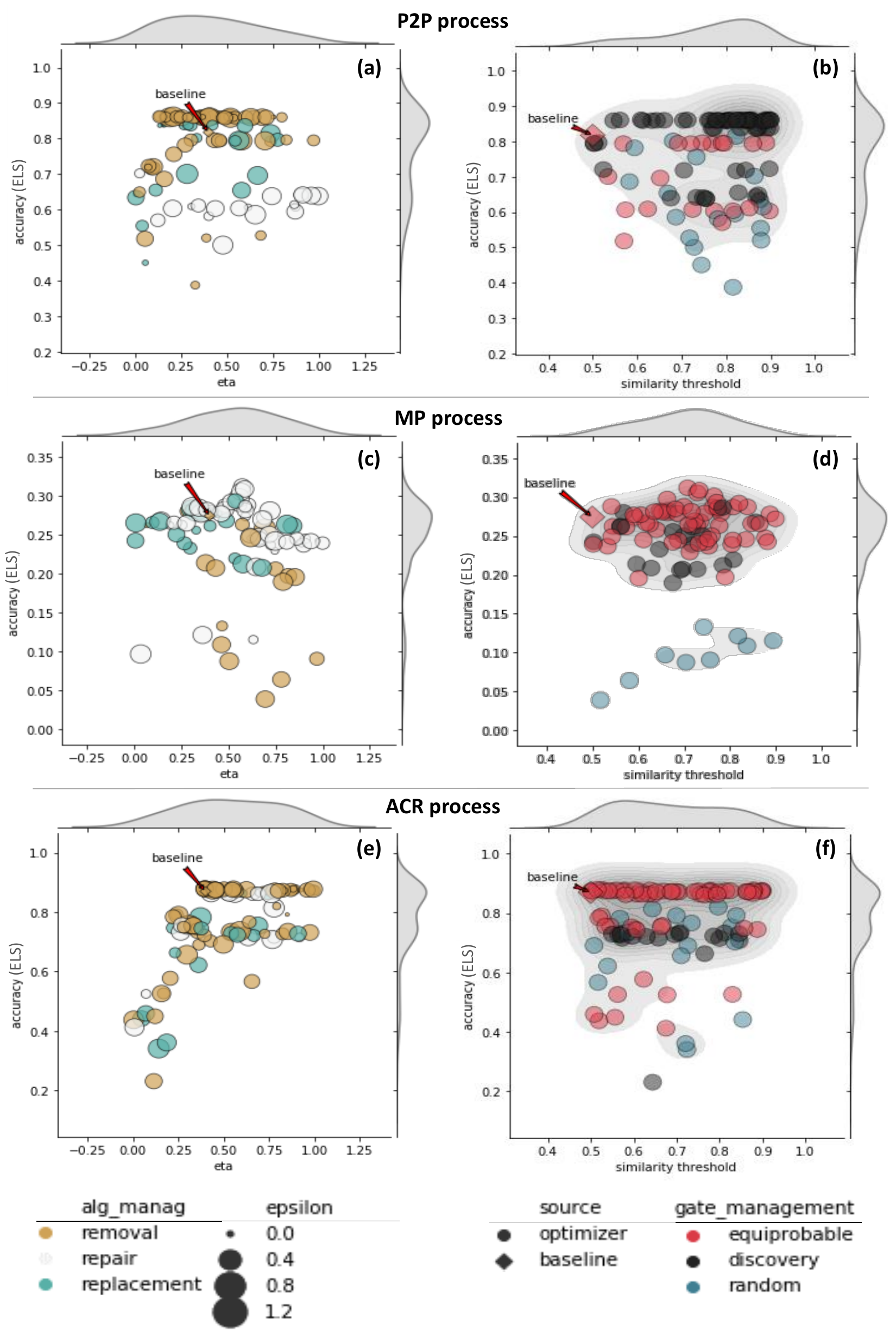}
    \caption{Hyper-parameter optimizer execution vs. baseline scenario results}
    \label{fig:scenarios}    
    \vspace*{-3mm}
\end{figure}

Figure~\ref{fig:scenarios} shows the accuracy results when simulating the baseline model and the BPS models generated by the optimizer for each of the event logs. Since the search space has six dimensions, we organize the results by grouping the hyperparamters according to the stage to which they belong, i.e., preprocessing or processing. Sub-figures \ref{fig:scenarios}a, \ref{fig:scenarios}c and \ref{fig:scenarios}e contrast the accuracy of the BPS models with the eta and epsilon values used for discoverying the process model and with the non-conformances handling technique. Sub-figures~\ref{fig:scenarios}b,~\ref{fig:scenarios}d, and~\ref{fig:scenarios}f plots the accuracy against the similarity threshold for discovering resource pools and against the method for determining the conditional branching probabilities.

In the P2P event log, we observe a linear association between using the Removal technique and model accuracy (see Figure~\ref{fig:scenarios}a). In the processing stage (see Figure~\ref{fig:scenarios}b), the branching probability discovery technique, in conjunction with similarity threshold values of around 0.85, have a strong linear association with accuracy. We also observe that the hyperparameter-optimized method clearly outperforms the baseline scenario. 

The MP log led to different results. In the the preprocessing stage (see Figure~\ref{fig:scenarios}c) we observe an association between using the Repair technique and higher accuracy values. In the processing stage (see Figure~\ref{fig:scenarios}d), there is a weak positive association between using the Equiprobable allocation of probabilities and model accuracy. The similarity threshold does not appear to be a determining factor. Compared to the baseline model, the optimizer could find better configurations; however, for this log, the choice of non-conformance handling technique played a more critical role.


In the ACR event log, in the preprocessing stage (see Figure~\ref{fig:scenarios}e), the Removal technique and the Repair technique have a strong association with higher model accuracy values, especially with eta values greater than 0.4. However, none of these two non-conformance handling techniques shows clear superiority. On the other hand and surprisingly, the Equiprobable allocation of branching probabilities leads to higher accuracy (see Figure~\ref{fig:scenarios}f). As was the case in the MP log, the similarity threshold does not appear to be a determining factor. For this log, the baseline method led to slightly lower accuracy than the hyperparameter-optimized method.

As shown in Table~\ref{tab:test}, the Mann-Whitney U test found that the differences in accuracy between the simulation runs of the optimized BPS model and the runs of the baseline BPS model are statistically significant for each of the three logs (i.e., null hypothesis rejected with p-values$<0.5$).

\begin{table}[ht]\scriptsize
\centering
\resizebox{0.8\textwidth}{!}{%
\begin{tabular}{@{}l|l|l|l@{}}
\toprule
\textbf{log} & \textbf{Baseline (ELS)} & \textbf{Optimizer (ELS)} & \textbf{p-values} \\ \midrule
P2P & 0.817526601 & 0.862158746 & 9.13E-05 \\ \midrule
MP & 0.274136518 & 0.311775719 & 9.13E-05 \\ \midrule
ACR & 0.869907699 & 0.880039013 & 0.018817657 \\ \bottomrule
\end{tabular}%
}
\caption{One tail Mann-Whitney U test results}
\label{tab:test}
\vspace*{-3mm}
\end{table}


\section{Threats to validity}

The experimental evaluation is restricted to one synthetic and two real-life event logs. As such, the generalizability of the results is limited: The results might be different for other event logs, and as shown in the evaluation, particularly for event logs for which the automated process discovery technique does not manage to discover an accurate process model.

Each parameter is extracted using a particular algorithm, because our focus was on automatic discovery of simulation models and the search for greater precision in relation to the process model used as the basis. One possible extension of this tool could include multiple extraction options for each parameter.

Due to the sequential nature of the BPS techniques, multitasking, batching and deliberate delays within a process due to relative priorities (e.g. a process being ``low priority''), are not taken into account in the proposed approach. Addressing these problems would require the development of new simulation techniques that are beyond the scope of this work.
\vspace*{-3mm}

%
\section{Conclusion and Future Work}
\label{sec:conclusion}

This paper outlined a method for automated discovery of business process simulation models from event logs and defined a measure for assessing the accuracy of a BPS model relative to an event log. The proposed method takes as input an event log, automatically discovers a process model, aligns the log to the model (and repairs it accordingly), and applies a range of replay and organizational mining techniques to extract all the parameters required for simulation. Once a BPS model is discovered, its accuracy is measured using a timed string-edit distance between the simulation log(s) it generates and the original log. A hyper-parameter optimizer is then used to search through the space of possible configurations so as to maximize the accuracy of the final BPS model. 

The proposed method has been implemented as an open-source tool, namely Simod, and evaluated using three real-life event logs from different domains. The evaluation shows that the hyper-parameter optimization method significantly improves the accuracy of the resulting BPS model, relative to an approach where default parameters are used. Also, it was observed that the best configuration found varies from one event log to another, further emphasizing the need for automated hyper-parameter optimization in this setting.


The evaluation reported in this paper is limited in terms of number of datasets due to the difficulty in obtaining access to real-life event logs where every (human) activity has both a start and an end timestamp, which is essential in order to determine the processing times of the activities. A direction for future work is to conduct a more systematic evaluation using a larger set of event logs, so as to identify possible relations between the characteristics of an event log and the associated (optimal) hyper-parameter settings and thus derive guidance for the configuration of BPS models.

Another limitation of the present study is that it relies on a relatively simple approach to business process simulation, in which activity instances (a.k.a.\ work items) are assigned to resources on a first-in, first-out basis (no notion of prioritization), each resource only performs one work item at a time (no multi-tasking), and a resource starts a work item assigned to it immediately after the assignment and works on it uninterruptedly until it is completed (no delays due to fatigue effects, no pauses, and no batching). Extending the proposed approach to lift these limitations and studying the effects of these phenomena on the accuracy of BPS models is another avenue for future work,



\medskip\noindent\textbf{Acknowledgments.} This research is funded by the European Research Council (PIX Project).

%

\bibliography{bib/references}

\end{document}